\documentclass{aipproc}
\layoutstyle{6s}

\newcommand{\nco}{\newcommand}
\nco{\beq}{\begin{equation}}
\nco{\eeq}{\end{equation}}
\nco{\beqa}{\begin{eqnarray}}
\nco{\eeqa}{\end{eqnarray}}
\nco{\lra}{\leftrightarrow}
\def\sfrac#1#2{{\textstyle{#1\over #2}}}
\def\eps{\epsilon}
\nco{\sss}{\scriptscriptstyle}
{\nco{\lsim}{\mbox{\raisebox{-.6ex}{~$\stackrel{<}{\sim}$~}}}
{\nco{\gsim}{\mbox{\raisebox{-.6ex}{~$\stackrel{>}{\sim}$~}}}

\begin{document}
\title{Order $\rho^2$ Corrections to Cosmology with Two Branes}

\author{J\'er\'emie Vinet$^\dagger$}{address={presenter },email={vinetj@physics.mcgill.ca},}
\author{James M. Cline}{address={McGill University, Montr\'eal, Qu\'ebec, Canada. },email={jcline@physics.mcgill.ca},}

\begin{abstract}

We present the results of work in which we derived the O($\rho^2$)
corrections to the Friedmann equations in the Randall-Sundrum I model.  The
effects of Golberger-Wise stabilization are taken into account.  We
surprisingly find that in the cases of inflation and radiation domination,
the leading corrections on a given brane  come exclusively from the effects
of energy density located on the opposite brane. \end{abstract}

\maketitle

\section{Introduction} In recent years, there has been considerable
interest in the idea that our world might be a 3-brane embedded in a higher
dimensional bulk \cite{ADD,RSI,RSII}.   The fact that gravity alone is
allowed to propagate through the bulk not only accounts for why the extra
dimension(s) have so far avoided detection, but also provides an attractive
solution to the infamous hierarchy problem.  

We will be concerned here mainly with the Randall-Sundrum I model, where
two 3-branes bound a slice of AdS(5) space.  In this model, the weakness of
gravity on our brane, the so-called ``TeV brane'' which has negative tension, is a
consequence of the fact that the extra dimension is warped, {\it i.e.,} the metric
along different slices of the bulk parallel to the 3-branes will depend on
their position  along the extra dimension.  

An interesting feature of models with 3-branes in a 5-dimensional bulk is
the fact that the Friedmann equations contain terms of higher order in the
energy density than in  standard cosmology\ \cite{BDL,CGKT,CGS,KKOP,CGRT}. 
More specifically, $H^2 = \frac{8\pi G}{3}\rho + O(\rho^2)$.   These new
terms could have important implications during inflation\
\cite{inf,MWBH,CLL} and  electroweak baryogenesis\ \cite{Geraldine, CGK}.  
Until now however, their exact form had only been worked out in the
Randall-Sundrum II model, which doesn't solve the hierarchy problem. (In RS
II, the extra dimension is  infinitely large, and there is only a single
3-brane present).  

We present here the results of work \cite{CV} (see also \cite{CF}) whose
aim was to find the O($\rho^2$) corrections to the Hubble rate in the RS I
model, taking into account the effect of the Goldberger-Wise (GW)
mechanism \cite{GW2} for stabilizing the extra dimension.  

\section{Einstein equations}
The Einstein equations follow from the action
\beqa
S &=& \int d^{\,5}x\sqrt{g} \left( -{1\over 2\kappa^2}R - \Lambda + 
    \sfrac{1}{2}\partial_{m}\Phi\partial^{m}\Phi
        -V(\Phi)\right)\nonumber\\
        &+&\int d^{\,4}x \sqrt{g}\left({\cal L}_{m,0} -
V_{0}(\Phi)\right)|_{y=0}
        +\int d^{\,4}x \sqrt{g}\left({\cal L}_{m,1} -
        V_{1}(\Phi)\right)|_{y=1},
\eeqa
where $\kappa^2=M_5^{-3}$, and the potential $V(\Phi)$ is left unspecified for now.
The brane contributions are a sum of matter, represented by 
\beq
	{\cal L}_{m,0}\sim\rho_*;\qquad  {\cal L}_{m,1}\sim\rho
\eeq
and tension, which is the value of the brane's scalar
potential $V_i(\Phi(y_i))$.  The matter Lagrangians cannot be written
explicitly for cosmological fluids, but their effect on the Einstein
equations  is specified through
their stress-energy tensors (\ref{stress}).
Our ansatz for the 5-D metric has the form
\beqa
\label{ansatz}
	ds^{2} &=& n^2(t,y)\,dt^{2} 
        -a^2(t,y)\sum_i dx_i^2 -b^{2}(t,y)\,dy^{2}
	\nonumber \\
               &\equiv& e^{-2N(t,y)}dt^2-
		a_0(t)^{2}e^{-2A(t,y)}
		\sum_i dx_i^2 -b^2(t,y)\,dy^{2}.
\eeqa
We will make a perturbative expansion in the energy densities
$\rho,\rho_*$ of
the branes around the static solution, where $\rho=\rho_*=0$:
\beqa
		N(t,y) &=& A_0(y) + \delta N_1(t,y) + \delta N_2(t,y); \qquad
		\!\!\!
		A(t,y) = A_0(y) + \delta A_1(t,y) + \delta A_2(t,y) \nonumber\\
		b(t,y) &=& b_0 + \delta b_1(t,y)+ \delta b_2(t,y) ; \qquad\qquad
		\Phi(t,y) = \Phi_0(y) + \delta\Phi_1(t,y) + \delta\Phi_2(t,y).
		\nonumber\\
\eeqa
The subscripts on the perturbations indicate their order in powers of 
$\rho$ or $\rho_{*}$.
This ansatz is to be substituted into the scalar field equation
and into the Einstein equations, $G_{mn}= \kappa^2 T_{mn}$.  (Since
the scalar field equation can be derived from a combination
of the Einstein equations, we will not worry about it any further).
The stress energy tensor is
$T_{mn} = g_{mn}(V(\Phi)+\Lambda)+\partial_{m}\Phi\partial_{n}\Phi
-\frac{1}{2}\partial^{l}\Phi\partial_{l}\Phi g_{mn}$ in the bulk.
On the branes, $T_m^{\ n}$ is given by
\begin{eqnarray}
T_{m}^{\ n} &=&{\delta(y)\over b(t,0)} \,{\rm 
diag}(V_{0}+\rho_{*},V_{0}-p_{*},V_{0}-p_{*},V_{0}-p_{*},0)
\nonumber \\  \label{stress}
&+&{\delta(y-1)\over b(t,1)}\,{\rm diag}(V_{1}+\rho,V_{1}-p,V_{1}-p,V_{1}-p,0)
\end{eqnarray}
(Later we will assume that the potentials $V_0$ and $V_1$ are very stiff
and are vanishing at their minima, so they can be neglected.)

It will be useful to define the following variables, which appear naturally in the Israel
junction conditions:
\beqa
	\Psi_2 = \delta A'_2 - A'_0 {\delta b_2\over b_0}-\frac{\kappa^2}{3} \Phi_0' \delta \Phi_2-\frac{\kappa^2}{6}
	\left(\delta \Phi_1'+\Phi_0' 
	{\delta b_1\over b_0}\right) \delta \Phi_1;\qquad
	\Upsilon_2 = \delta N'_2 - \delta A'_2.
\eeqa
\beqa
\label{Psibc}
\Psi_2(t,0) &=&
 +\left.\frac{\kappa^{2}}{6}b_{0}\rho_{*} {\delta b_1\over
b_0}\right|_{t;\,y=0} ;\qquad\qquad
\Psi_2(t,1)=-\left.\frac{\kappa^{2}}{6}b_{0}\rho{\delta b_1\over b_0}
\right|_{t;\,y=1}\\
\label{Upsbc}
 \Upsilon_2(t,0) &=&
 -\left.\frac{\kappa^{2}}{2}b_{0}(\rho_{*}+p_*){\delta
b_1\over b_0}\right|_{t;\,y=0};\quad 
\Upsilon_2(t,1)= 
+\left.\frac{\kappa^{2}}{2}b_{0}(\rho+p){\delta b_1\over
b_0}\right|_{t;\,y=1}
\eeqa
The analogous quantities at first order were found to be
\beq
\label{PsiUpseqs}
	\Psi_1 = \delta A' - A'_0 {\delta b_1\over b_0}-\frac{\kappa^2}{3} 
	\Phi_0' \delta\Phi_1;\qquad
	\Upsilon_1 = \delta N'_1 - \delta A'_1
\eeq
in \cite{CF}, and they satisfy the same boundary conditions as in
(\ref{Psibc},\ref{Upsbc}), but with the replacement ${\delta b_1\over b_0}\to 1$.

In terms of these variables, we can write the second order Einstein 
equations as
\beqa
\label{00eq} 
	 \left({\dot a_0\over a_0}\right)_{\!\!(2)}^2 b_0^2 e^{2A_0} &=&
4 A_0' \Psi_2 - \Psi'_2 +{\cal F}_{\Psi} \\
\label{00iieq} 
	2\left( \left({\dot a_0\over a_0}\right)_{\!\!(2)}^2 -
 	\left({\ddot a_0\over a_0}\right)_{\!\!(2)}\right) 	b_0^2 e^{2A_0} 
	&=&- 4 A_0'\Upsilon_2  + \Upsilon'_2+{\cal F}_{\Upsilon} \\
\label{05eq} 
	0 &=& - \left({\dot a_0\over
	a_0}\right)_{\!\!\!\left(\sfrac12\right)}\!\!\Upsilon_2 + \dot\Psi_2
	+{\cal F}_{05}\\
\label{55eq} 
\left( \left({\dot a_0\over a_0}\right)_{\!\!(2)}^2 +
       \left({\ddot a_0\over a_0}\right)_{\!\!(2)}\right)b_0^2 e^{2A_0} &=&
	A_0'(4\Psi_2 + \Upsilon_2) + \frac{\kappa^2}{3}\left( \Phi_0''\delta
	\Phi_2-\Phi_0'\delta
	\Phi_2'+\Phi_0'^{2}\frac{\delta b_2}{b_{0}}\right) \nonumber\\
       && \qquad\qquad\qquad + {\cal F}_{55}	
\eeqa
where all the dependence on first order quantities squared is contained in the
functions ${\cal F}_{\Psi}$, ${\cal F}_{\Upsilon}$, ${\cal F}_{05}$ and 
${\cal F}_{55}$.  Here, the equivalent first order equations
can be obtained by simply replacing the subscripts $2$ for $1$, and leaving out
the functions ${\cal F}$.

We thus have at each order in the perturbative expansion a set of first order differential
equations which, when combined with the boundary conditions, allows us to solve for the
unknown functions $\Psi_n$, $\Upsilon_n$ and $\delta b_n / b_0$.  (We will work in 
a gauge where the fluctuations $\delta \Phi_n$ vanish).  

One final note regarding the physical value of the Hubble rate.  Since the (00) component of the metric receives corrections when carrying out our perturbative
expansion, we should use the time coordinate
\beqa
\label{taueq}
	d\tau = {n(t,1)\over n_0(t,1)}\, dt = e^{-\delta N_1(t,1) -
	\delta N_2(t,1) - \cdots}\, dt
\eeqa
to define the Hubble rate.  This means that physical Hubble rate is given by
\beqa
\label{dNcorr}
	H \cong (1 + \delta N_1)
	{\dot a\over a} 
\eeqa
rather than simply $H={\dot a\over a}$ as we might naively have expected.  (See \cite{CV} for a more thourough discussion of this issue.)

\section{Friedmann equations}
Taking into account all that has been said in the previous section, we are now able to give
the Friedmann equations to second order in $\rho$ and $\rho_*$:
\beqa
\label{H21}
H^2|_{y=1} &=& {8\pi G\over 3} \left( \bar\rho + \rho_* +
	{2\pi G\over 3 m^2_r\Omega^2} \left( 
9(1-3\omega)(1+\omega)\bar\rho^2  \right.\right.\nonumber\\
 && \left.\phantom{8\pi G\over 3}\left. +
4(1-3\omega_*)(4+3\omega_*)\Omega^2\rho_*^2
+4(1-3\omega)(4+3\omega)\bar\rho\rho_* ) \right)\right)\\
\label{H20}
H^2|_{y=0} &=&   {8\pi G\over 3} \left( \bar\rho + \rho_* +
	{2\pi G \over 3 m^2_r\Omega^2} \left( 
9(1-3\omega_*)(1+\omega_*)\rho_*^2\Omega^4  -
(1-3\omega)(7+3\omega)\bar\rho^2
\right.\right. \nonumber\\
 && \left.\phantom{8\pi G\over 3}\left. 
+ 2\Omega^{2}\bar\rho\rho_*[ 2(1-3\omega)(2+3\omega) - 3(1-3\omega_*)(1+\omega) 
 ] \ \right)\right)\\
\label{dH1}
\left.{dH\over d\tau}\right|_{y=1} &=&-4 \pi G \left(\bar\rho (1+\omega) 
+ \rho_* (1+\omega_*) \phantom{1\over 3}\right.\nonumber\\
  &+& \left.\phantom{-4 \pi G\!\!\!\!\!\!\!\!\!\!\!\!\!\!\!\!}
{4 \pi G \over 3 m_r^2 \Omega^2}
\left(\Omega^2(1+\omega_*)(1-3\omega_*)(13+9\omega_*)
\rho_*^2+9(1-3\omega)(1+\omega)^2\bar\rho^2
\phantom{1\over 3}\!\!\!\!\!\right. \right. \nonumber\\
 &-&\left.\left. \phantom{-4 \pi G\!\!\!\!\!\!\!\!\!\!\!\!\!\!\!\!}
(1-3\omega)(2(1+\omega_*)
+2(1+\omega)+6(1+\omega)^2+3(1+\omega)(1+\omega_*))
{\bar\rho}\rho_* \right)\phantom{1\over 3}\!\!\!\!\!
\right)\nonumber\\
\\
\label{dH0}
\left.{dH\over d\tau}\right|_{y=0} &=&-4 \pi G 
\left(\bar\rho (1+\omega) + \rho_* (1+\omega_*)\phantom{1\over 3} \right.\nonumber\\
 &+&\left. \phantom{-4 \pi G\!\!\!\!\!\!\!\!\!\!\!\!\!\!\!\!} 
{4 \pi G \over 3 m_r^2 \Omega^2}\left(-4(1+\omega)
(1-3\omega)\bar\rho^2+9(1+\omega_*)^2(1-3\omega_*)\Omega^4 \rho_*^2
\phantom{1\over 3}\!\!\!\!\!\right.\right. \nonumber\\
&+&\left.\left.\phantom{-4 \pi G\!\!\!\!\!\!\!\!\!\!\!\!\!\!\!\!}
\Omega^2\left[6(1-3\omega)(1+\omega)^2+2(1-3\omega)(1+\omega)-2(1-3\omega)
(1+\omega_*)\right.\right.\right.\nonumber\\
&-&\left.\left.\left.\phantom{-4 \pi G\!\!\!\!\!\!\!\!\!\!\!\!\!\!\!\!} 
4(1-3\omega_*)(1+\omega)+3(1+\omega_*)(1-3\omega)(1+\omega)
\phantom{\Omega^2\!\!\!\!\!\!\!}\right]
\rho_* {\bar\rho}\phantom{1\over 3}\!\!\!\right)\right)
\eeqa
where we have used
\beqa
\dot\rho &=& -3H(\rho + p) \equiv -3H(1+\omega)\rho;\quad
\label{8piG}
	8\pi G =\kappa^2\left(2b_0
	\int_0^1e^{-2A_0}dy\right)^{-1};\\
\dot\rho_*&=& -3H(\rho_* + p_*) \equiv -3H(1+\omega_*)\rho_*;\quad
\label{radmass}
 m_r^2\cong\frac{4}{3}\kappa^2 v_0^2\eps^2 k^2\Omega^{2+2\eps};\\
\bar\rho &=& \Omega^4 \rho ;\qquad\qquad\qquad
\Omega = e^{-A_0(y=1)}.
\eeqa
It should also be noted that in order to obtain analytical results, we had
to expand our expressions in powers of $\Omega$, and we have only kept the
dominant terms.

The careful reader will notice that the second order corrections vanish identically for 
radiation-like equations of state ($\omega = \omega_* = 1/3$).  In that case, we need to 
look at next to leading order terms to find non-vanishing corrections:
\beqa
  H^2|_{y=1} &=& {8\pi G\over 3}\left((\rho_*+\bar\rho) +
\frac{2 \pi G}{3k^2 \Omega^4} 
	(\Omega^2\rho_*-\bar\rho)\rho_*\right)\\
H^2|_{y=0} &=& {8\pi G\over 3}\left((\rho_*+\bar\rho) -
\frac{2 \pi G}{3k^2 \Omega^6} 
	(\Omega^2\rho_*-\bar\rho)\bar\rho\right)
\eeqa
and the equations for $dH/d\tau$ have vanishing corrections at this order in
$\rho$ and $\rho_*$.
\section{Discussion}
Let's now look at what our results mean. First, we note that the brane on which 
the hierarchy problem is solved is the one located at $y=1$.  It is thus natural 
to assume this is the brane we are living on.  If we look strictly at the terms 
linear in $\rho, \rho_*$, we see that in order for the current Hubble rate on our brane 
not to be completely dominated by the energy density on the other brane, we must 
assume that $\rho_*$ is presently very small or vanishing.  

If we now look at the second order terms on our brane during inflation 
($\omega = -1$), we see that the only corrections come from terms involving $\rho_*$.  
So unless we can come up with a model where $\rho_*$ was large in the past and becomes 
negligible in the present epoch, there can have been no appreciable effect coming from 
the second order terms during inflation.

The situation during radiation domination is similar.  The leading second order corrections
on each brane involve the other brane's energy density.  Furthermore, even if for the sake 
of argument we allow $\rho_*$ to have been present during this era, we know that it can't
be more than $10\%$ of the value of $\bar\rho$ by the time of nucleosynthesis.  This means
that we can ignore the $\Omega^2\rho_*$ term, and that the term that is left over suppresses
the Hubble rate.  But in order to make the sphalerons go out of equilibrium at the electroweak
phase transition, we would have needed to make the Hubble rate larger \cite{Geraldine}.  So 
once again, we find that the second order corrections do not seem to have
any constructive applications in this model.

\section{Conclusion}

We have derived the second order corrections to the Friedmann equations in
the Randall-Sundrum I  model, taking into account the effect of the GW
stabilization mechanism.  We have found that the  corrections during
radiation domination have the wrong sign to be useful for electroweak
baryogenesis.  We have also found that any effect on a brane with an
inflationary equation of state can only be coming from  the opposite
brane.  Our approach being perturbative in nature, we must however keep in
mind that our results should not be expected to hold when the energy
density becomes greater than some critical density, which on the TeV
brane is the (TeV)$^4$ scale.

\subsection{Acknowledgements}
We wish to thank Hassan Firouzjahi for enlightening discussions.  
J.V.\ is supported in part by a grant from Canada's NSERC.

\end{document}